\documentclass[twocolumn,showpacs,preprintnumbers,amsmath,amssymb,floatfix,superscriptaddress]{revtex4-1}
\usepackage{graphicx}
\usepackage{epsfig}
\usepackage{amsfonts}
\usepackage{amssymb}
\usepackage{epsf}
\newcommand{\insertplot}[5]{\begin{figure}
 \hfill\hbox to 0.05in{\vbox to #5in{\vfill
 \inputplot{#1}{#4}{#5}}\hfill}
 \hfill\vspace{-.1in}
 \caption{#2}\label{#3}
 \end{figure}}
\newcommand{\inputplot}[3]{
 \special{ps: plotfile #1}
}

\newcounter{fig}

\begin{document}

\title{
Traversable wormholes in  Einstein-Dirac-Maxwell theory}

\author{{\bf Jose Luis Bl\'azquez-Salcedo}}
\affiliation{ Departamento de F\'isica Te\'orica and IPARCOS, Universidad Complutense de Madrid, E-28040 Madrid, Spain}
\affiliation{ Institut f\"ur  Physik, Universit\"at Oldenburg, Postfach 2503,
	D-26111 Oldenburg, Germany}

\author{ {\bf Christian Knoll}}
\affiliation{ Institut f\"ur  Physik, Universit\"at Oldenburg, Postfach 2503,
	D-26111 Oldenburg, Germany}

\author{{\bf Eugen Radu}}
\affiliation{
{Departamento de Matem\'atica da Universidade de Aveiro and } 
\\ 
{\small  Centre for Research and Development  in Mathematics and Applications (CIDMA),} 
\\ 
{\small    Campus de Santiago, 3810-183 Aveiro, Portugal}
}
\date{\today}
\begin{abstract}
We construct a specific example of a class of traversable wormholes in Einstein-Dirac-Maxwell
theory in four spacetime dimensions,
without needing any form of exotic matter.
Restricting to a model with two massive fermions
in a singlet spinor state,
we show the existence of spherically symmetric asymptotically flat configurations which  
are free of singularities,
representing localized states.
These solutions satisfy a generalized Smarr relation,
being connected with the extremal Reissner-Nordstr\"om black holes.
They also possess a finite mass $M$ and electric charge $Q_e$,
with $Q_e/M>1$. 
An exact  wormhole solution with ungauged,  massless fermions is also reported.
\end{abstract}
\maketitle

\noindent{\textbf{~~~Introduction.--~}}
The wormholes (WHs) have entered modern physics soon after the discovery of 
black holes (BHs)
\cite{Flamm,Einstein-Rosen}. 
In both cases it took decades to understand their rich physical
content and to realize that they may play a role in Nature.
However, while there is increasing evidence for the existence of (astrophysical)  BHs,
the (Lorentzian, traversable) WHs remain so far 
rather an interesting possibility, {although with observational implications \cite{Wielgus:2020uqz}}.
A basic difference between these two types of solutions 
occurs already at the level of
energy-momentum supporting the corresponding geometries.
 While the BHs exist in vacuum, being the end point 
of (normal matter's) gravitational collapse, 
the traversable  WHs
necessarily require a matter content violating the null energy condition 
\cite{Morris:1988cz},
\cite{Visser}.
Restricting to  a field theory source 
and
a classical setting, the (bosonic) matter fields necessarily possess a non-standard Lagrangian
($e.g.$ 'phantom' fields 
\cite{phantom}),
or one has to consider extensions of gravity beyond general relativity
(see $e.g.$
\cite{Kanti:2011jz},
\cite{Barcelo:2000zf}).

However, as we shall prove in this work,
the situation changes for fermions, 
with the existence of traversable WHs solutions of the Einstein-Dirac equations.
In our approach,
the Dirac matter is described by a quantum wave function rather than a quantum field.
This results in  a more  tractable model,
with the backreaction of the matter to
 spacetime geometry being taken into account.
Moreover,  the inclusion of an electric charge leads to 
`smooth' geometries, without the presence of a thin shell of extra-matter
at the throat of the WH.

\noindent{\textbf{~~~ Einstein-Dirac-Maxwell model.--~}}
We consider a model with two gauged {relativistic} fermions,  the spin of which is taken 
to be opposite in order to satisfy spherical symmetry.
Working in units with 
$G=c=\hbar=1$,
the action of the corresponding Einstein-Dirac-Maxwell (EDM) model reads 
\begin{eqnarray}
	\label{action}
	S = \frac{1}{4\pi}
\int \mathrm d^4 x \sqrt{-g} \, 
	\left[ \frac{1}{4}R+ \mathcal L_D-\frac{1}{4}F^2    \right] \, ,
\end{eqnarray}
where $R$ is the Ricci scalar of the metric 
$g_{\mu\nu}$, 
$F_{\mu\nu}=\partial_{\mu}A_\nu-\partial_{\nu}A_\mu$
 is the field strength tensor of the U(1) field $A_\mu$,
 and
\begin{eqnarray}
\label{Lagrangian_Dirac}
\nonumber
\mathcal L_D = 
\sum\limits_{\boldsymbol{\epsilon}=1,2} 
\left[ \frac{\mathrm i}{2} \overline{\Psi}_{\boldsymbol{\epsilon}} 
\gamma^\nu \hat{D}_\nu \Psi_{\boldsymbol{\epsilon}} 
- \frac{\mathrm i}{2} 
\hat{D}_\nu \overline{\Psi}_{\boldsymbol{\epsilon}}  \gamma^\nu \Psi_{\boldsymbol{\epsilon}}  
- \mu 
\overline{\Psi}_{\boldsymbol{\epsilon}}  \Psi_{\boldsymbol{\epsilon}}  
\right] , 
\end{eqnarray}
where $\gamma^\nu$
are the curved
space gamma matrices   \cite{note1}
and $\mu$ is the mass of {\it both} spinors $\Psi_{\boldsymbol{\epsilon}=1,2}$.
Also, 
$
\hat{D}_\mu  =  \partial_{ \mu} + 
 \Gamma_\mu -
	i q A_\mu 
$,
where $\Gamma_\mu$ are the spinor connection matrices,
and $q$ is the gauge coupling constant. 
  The resulting field equations are  
\begin{eqnarray}
\label{Einstein}
&&
R_{\mu \nu} -\frac{1}{2}R g_{\mu \nu}= 2T _{\mu \nu}~{\rm with}~T _{\mu \nu}=T^{(D)}_{\mu \nu}+ T^{(M)}_{\mu \nu},~~
\\
\label{matter}
&&
(\gamma^\nu \hat{D}_\nu-\mu) \Psi_{\epsilon} =0,~~\nabla_{\mu}F^{\mu \nu}=q  j^\nu,
\end{eqnarray}
with
the current
 $j^{\nu}=\sum\limits_{\epsilon=1,2}\overline{\Psi}_{\epsilon} \gamma^{\nu} \Psi_{\boldsymbol{\epsilon}}
$
and
$
T^{(D)}_{\mu \nu} = \sum\limits_{\boldsymbol{\epsilon}=1,2} 2\Im ( \overline{\Psi}_{\boldsymbol{\epsilon}} \gamma_{(\mu} \hat{D}_{\nu)} \Psi_{\boldsymbol{\epsilon}} ),
$
$
T^{(M)}_{\mu\nu} =  F_{\mu\alpha} F_\nu^{\alpha} - \frac{1}{4} F^2 g_{\mu\nu}.
$

Restricting to static, spherically-symmetric
solutions of the field equations, we consider a general metric ansatz
$
ds^2=g_{tt}(r)dt^2+g_{rr}(r)dr^2+g_{\Omega \Omega}(r)d\Omega^2,
$
where 
$r$ and $t$ are the radial and time coordinates, 
and
$d\Omega^2=d\theta^2+\sin^2\theta d\varphi^2$.
The U(1) field is purely electric, with 
$
A=V(r) dt.
$
A general spinors Ansatz compatible with the symmetries of
the considered line element is \cite{Blazquez-Salcedo:2019uqq}
\begin{eqnarray}
\Psi_\epsilon = \mathrm e^{-\mathrm i w t}
{\cal R}_\epsilon(r)
\otimes \Theta_\epsilon(\theta,\varphi) \, ,
\end{eqnarray}
 with $w$ the frequency and
\begin{eqnarray}
&&
\nonumber
{\cal R}_1=-i{\cal R}_2=
 \bigg[ 
                          \begin{array}{c} 
\phi (r)
\\ 
-i \bar \phi(r) 
                          \end{array}
 \bigg] ~,
\\
&&
\nonumber
\Theta_1= \bigg[ 
                          \begin{array}{c} 
-\kappa \sin \frac{\theta}{2} 
\\ 
\cos \frac{\theta}{2} 
                          \end{array}
 \bigg] e^{i\frac{\varphi}{2}},~~
\Theta_2= \bigg[ 
                          \begin{array}{c} 
 \kappa \cos \frac{\theta}{2} 
\\ 
\sin \frac{\theta}{2} 
                          \end{array}
 \bigg] e^{-i\frac{\varphi}{2}},~~
\end{eqnarray}
with $\kappa=\pm 1$.
Also, assuming $r>0$, one considers 
the usual tetrad choice, with
 $e^r=\sqrt{g_{rr}}dr$, 
 $e^\theta=\sqrt{g_{\Omega \Omega}}d\theta$, 
 $e^\varphi=\sqrt{g_{\Omega \Omega}}\sin \theta d\varphi$,
 $e^t=\sqrt{-g_{tt}}dt$.

A useful parametrization in the numerics  is  
$\phi = |\phi|e^{i\alpha/2}= e^{i\pi/4} F-e^{-i\pi/4} G$.
Then the entire matter content of the model
is  encoded in the two real fermion functions $F(r), G(r)$,
together with the electrostatic potential $V(r)$.
This is essentially the framework used in 
\cite{Finster:1998ux} to construct (topologically trivial)
particle-like solutions of the EDM system.
In what follows we show that the system possess also 
traversable WH configurations \cite{note2}.

\noindent{\textbf{~~~ An exact solution.--~}} 
The resulting EDM equations can be solved  analytically in the $q=0$ limit, 
 the spinor fields being massless, with $w=0$.
The solution
has
 the metric and the U(1) potential
\begin{eqnarray}
&&ds^2=-(1-\frac{M}{r})^2dt^2
+\frac{dr^2}{(1-\frac{r_0}{r})(1-\frac{Q_e^2}{r_0r})}
+r^2 d\Omega^2, 
\\ 
\nonumber
&&V(r)=%
\frac{M}{Q_e}
\sqrt{(1-\frac{r_0}{r})(1-\frac{Q_e^2}{r_0r})},
~{\rm with}~~M=\frac{2Q_e^2 r_0}{Q_e^2+r_0^2}~,
\end{eqnarray}
while the spinor functions are 
\begin{eqnarray}
\nonumber
F(r)= c_0  { \bigg(\sqrt{1-\frac{Q_e^2}{r_0 r}}- \kappa \sqrt{1-\frac{r_0}{r} } \bigg)^2}/
{\sqrt{ 1-\frac{M}{r} } },~~
\\
\nonumber
G(r)=\frac{\kappa r_0}{32 c_0(Q_e^2+r_0^2)}
\frac{ \left(\sqrt{1-\frac{Q_e^2}{r_0 r}}+ \kappa \sqrt{1-\frac{r_0}{r} } \right)^2}
{\sqrt{ 1-\frac{M}{r} } },
\end{eqnarray}
with $c_0\neq 0$ an arbitrary constant.
This describes a (regular) traversable WH solution, with  $r_0$ the throat's radius and $Q_e<r_0$
the electric charge, while $M$
is 
the ADM mass (note that $Q_e/M>1$).
The WH geometry 
is supported by the spinors contribution to the total energy-momentum tensor,
being regular everywhere. 
 As 
$Q_e\to r_0$,
the extremal Reissner-Nordstr\"om (RN) BH is approached, while  
$T^{(D)}_{\mu\nu}\to 0$.

Although this solution captures some basic properties 
of the general configurations below, it also possesses some undesirable features. 
In particular, the spinor wave function is not normalizable, 
since $|\phi|$ does not vanish as $r\to \infty$.
However, the situation changes in a model with massive fermions,
they becoming exponentially localized.
%

%
\noindent{\textbf{~~~ The general case.--~}}The generic solutions have 
$\mu\neq 0$, 
$q \neq 0$
and are found numerically, by employing a metric ansatz
which makes transparent the WH structure and simplifies the numerics 
 \cite{Kanti:2011jz}
\begin{eqnarray}
\label{metric}
 ds^2=-e^{2\nu(r)}dt^2+f(r) dr^2+(r^2+r_0^2)d\Omega^2, 
\end{eqnarray}
with $r_0>0$
the radius of the throat, which is located at $r=0$
(with $A_T=4 \pi r_0^2$ the throat area).
The WH consists in two different regions 
$\Sigma_\pm$
of the same Universe.
The `up' region is found for $0<r<\infty$;
there is also a 'down' region, with $-\infty<r<0$. 
However, 
in general the joining at $r=0$
of these regions
is not `smooth', 
with a discontinuity of the metric derivatives.
This implies
 the presence of a thin mass shell structure at the throat,
with   a $\delta$-source added to the action (\ref{action})
($e.g.$ the surface energy density is $\epsilon_T=-4\nu'(0)/\sqrt{f(0)}$). 
The condition for a `smooth' geometry is
$\nu'(0)=0$.

Also, we shall consider the case of a symmetric  WH,
  the geometry (\ref{metric}) 
	and the energy-momentum tensor
	being invariant under the transformation $r\to -r$. 
The sign change of $r$  at the WH's throat  
reflects 
in a change of sign of the tetrad 
\cite{Cariglia:2018rhw}.
Then the matter functions transform  as
$V(r')=-V(r)$
and 
$\phi(r')= i \bar \phi(r)$
(with $r'=-r>0$),
while $\kappa \to -\kappa$ and $w\to -w$.
	As such, in what follows we shall report
	results  mainly for the
	$r \geq 0$ region.

With this framework, the problem reduced to solving 
 a system of four first order equations for 
$\{\nu,f,F,G\}$ and a second order equation for $V$ \cite{num}. 
These equations are invariant under the 
 transformation 
$w\to w+\beta$,
$V\to V+\beta/q$
(with $\beta$ an arbitrary constant),
which is fixed by imposing the electric potential to vanish at the throat.

The only global charges are the mass $M$ and the electric charge $Q_e$,
which are read from the far field asymptotics.
For the `up' region, one finds
	$\nu \to -M/r$,
$f \to 1+2M/r$,
$V \to \Phi-Q_e/r$,
(with $\Phi$ the electrostatic potential).
The spinor functions decay 
as 
$
$
$
 e^{-\mu_{*}r}/r,
$
where
$\mu_*=\sqrt{\mu^2-(w-\Phi/q)^2}$
(with the bound state condition $\mu_{*}^2 \geq 0$).

An approximate solution can also be found close the throat,
with the boundary conditions 
$\nu(0)=\nu_0$,
$f(0)=f_0$,
$F(0)=  0$,
$G(0)=G_0$
and 
$V(0)=0$
($\nu_0$, 
$f_0$,
$G_0$ 
being
nonzero constants).

The WHs satisfy a Smarr law,
the mass being the sum of an electrostatic term and a
 bulk contribution 
\begin{eqnarray}
M= \Phi Q_e+M_{(B)},
\end{eqnarray}
with
\begin{eqnarray}
\nonumber
&
M_{(B)}=4\int_0^{\infty}
dr (r^2+r_0^2)
\big(
\mu\sqrt{f}e^{\nu}(F^2-G^2)
+q V |\phi|^2\sqrt{f}
\big).
\end{eqnarray}
 By integrating the Maxwell equations,
one finds
	\begin{eqnarray}
	\label{relQ}
	Q_e=2 q Q_N+Q_{T},
	\end{eqnarray}
	where $Q_N$ is the Noether charge of a spinor {(or number of particles)}
	\begin{eqnarray}
\label{cond}
\nonumber
Q_{N}= \frac{1}{4\pi}\int_{\Sigma_+} d^3x \sqrt{-g}j^t_{\boldsymbol{\epsilon} } 
=4 \int_0^{\infty} dr \sqrt{f}(r^2+r_0^2) |\phi|^2,
\end{eqnarray}
and $Q_T=V'(0)e^{-\nu(0)}r_0^2/\sqrt{f(0)}$.
Similar relations hold for the $r<0$ region,
with mass, electric charge and Noether charge changing sign.
 
The equations of the model are invariant under the scaling 
transformation
(the variables and quantities
which are not specified remain invariant):
 $(r,r_0)\to \lambda (r,r_0)$,  
$(F,G) \to (F,G)/ \sqrt{\lambda}$,
$(\mu,q,w) \to (\mu,q,w)/\lambda$,
where $\lambda$ is a positive constant, 
while various quantities of interest transform as
$
(M,Q_e)\to \lambda (M,Q_e),~(Q_{N },A_H) \to \lambda^2 (Q_{N },A_H).
$
	Only quantities which are invariant under this transformation 
	(like 
	$M/Q_e$)
	are relevant.

As with the solitons 
\cite{Finster:1998ux},
\cite{Finster:1998ws},
\cite{Herdeiro:2017fhv},	
this transformation is used to impose the one particle condition,
$Q_N=1$,
for {\it each}  spinor in 
  both 
'up' or 'down' regions.
%

\begin{figure}[h!]
\centering
 \includegraphics[height=2.in,angle=0]{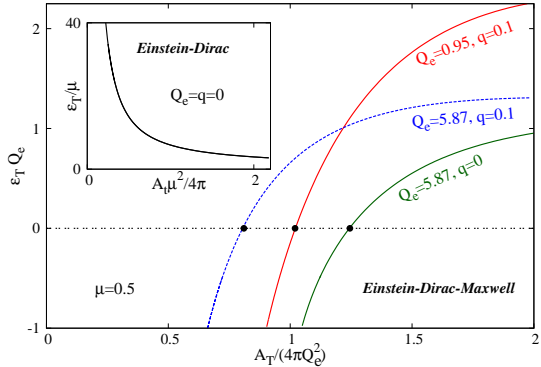} 
\caption{ The scaled  thin layer energy density at the throat, 
$\epsilon_T$, 
is shown as a function of the scaled 
throat area $A_T$ for several sets of solutions at fixed frequencies.
}
\label{junction}
\end{figure}

\begin{figure}[h!]
\centering
\includegraphics[height=2.in]{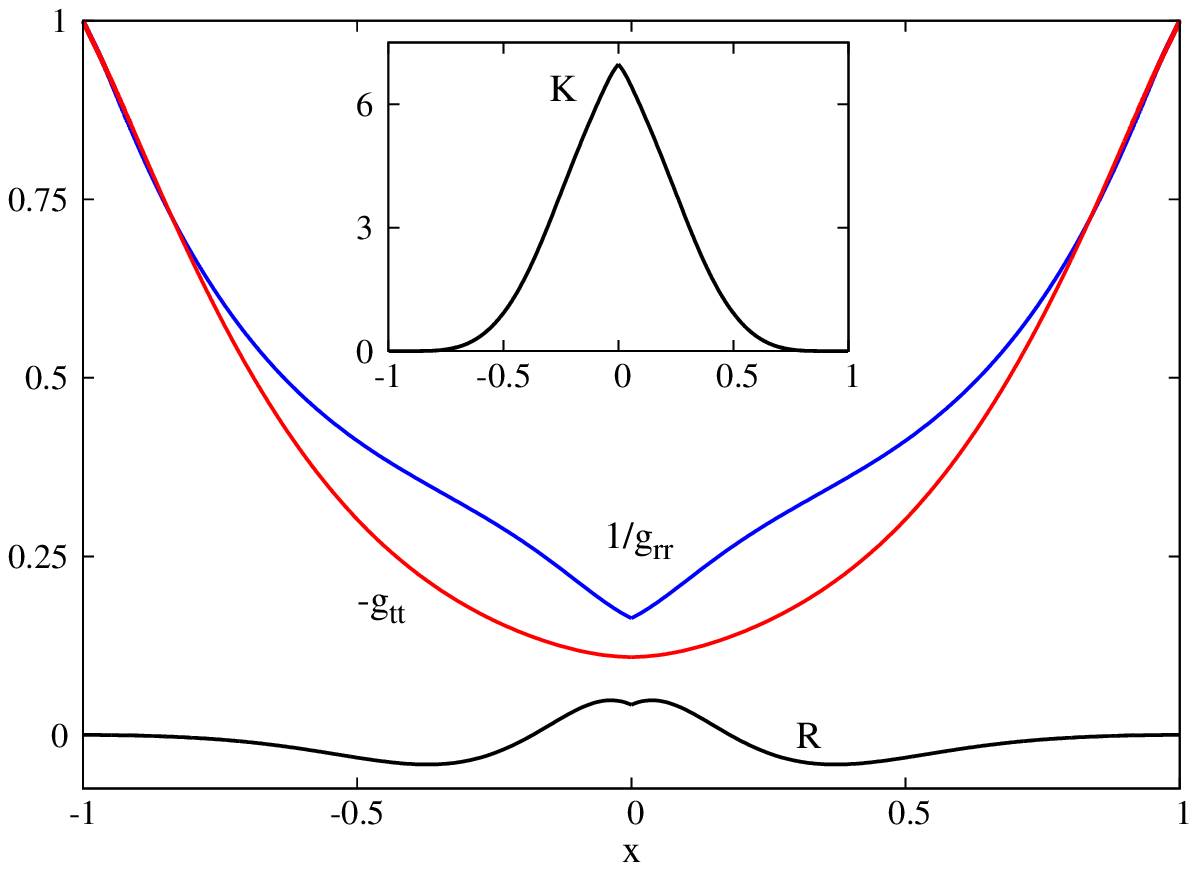} 
\includegraphics[height=2.in]{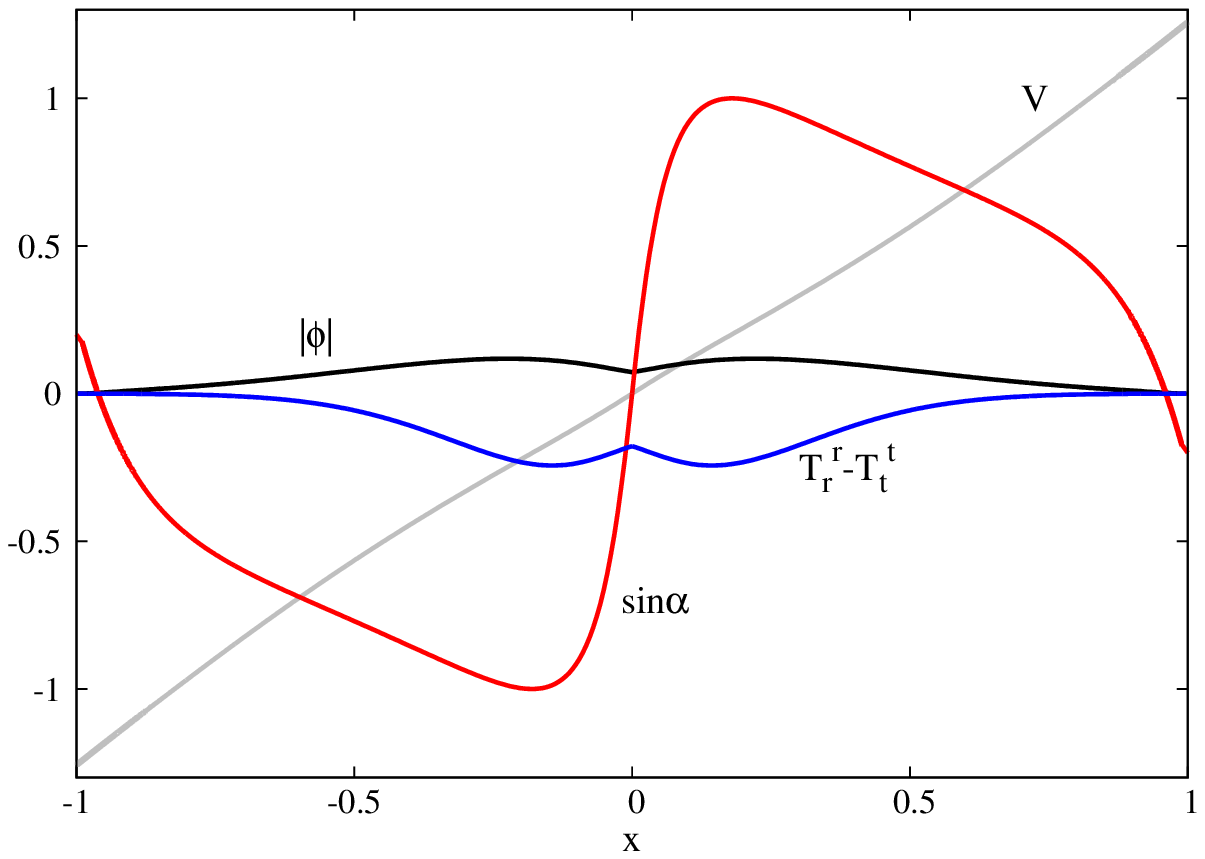}   
\caption{  
The profile of a typical wormhole solution is displayed together 
with the Ricci and Kretschmann scalars 
(with $x=\frac{2}{\pi}\arctan(r/r_0)$  a compactified radial coordinate). 
The violation of the null energy condition
is also shown.
}
\label{profile}
\end{figure}

\noindent{\textbf{~~~ The solutions.--~}}
We have solved the EDM for various values of the  model's constants 
$(\mu,q)$.
In particular, WH solutions
  exist also in the ED limit ($i.e.$ $q=0$ and $V=0$).
	However, as seen in the inlet of Fig. 1, those solutions have always $\nu'(0)\neq 0$,
and thus require the presence of extra-matter at the throat.
The `smooth' configurations necessarily possess a nonzero electric charge and have 
$\mu > 0$ (although $q$ can vanish),
the profile of a typical such configuration
(marked with a star in Fig. 3).
 being displayed in Fig. 2. 

In our approach,
apart from 
$(q,\mu)$,
the other input parameters are 
$\{Q_e,r_0,w\}$,
all other quantities ($e.g.$   $M$ and $Q_N$)
being read from the numerical output. 
As shown in Fig. 1, our results 
	indicate that 
 for fixed electric charge and field frequency, 
a  solution with no extra-matter at the throat
exists
	for a unique value of the throat size  \cite{note3}.

As such, when varying $w$, a continuous set of 'smooth' solutions is found, 
the corresponding picture
in terms of  mass $vs.$ throat area being shown in Fig. 3 
 (with the quantities given in units of the electric charge).
A curve  there  interpolates 
between the extremal RN BH (in which limit $r_0$ becomes
the horizon radius while the spinor fields vanish), 
and a critical configuration with  $\mu_{*} \to 0$.
This behaviour is generic, being found for
all considered values of $(q,\mu)$.
The set of all critical configurations 
forms the {\it critical line}. 
Although they still possesses  a 
smooth geometry, 
their ADM mass is negative, a feature shared by a set of solutions close to them.
Also, we have found that all solutions constructed so far 
have
$Q_e/M>1$ and
$q/\mu<1$.

\begin{figure}[h!]
\centering
\includegraphics[height=2.in,angle=0]{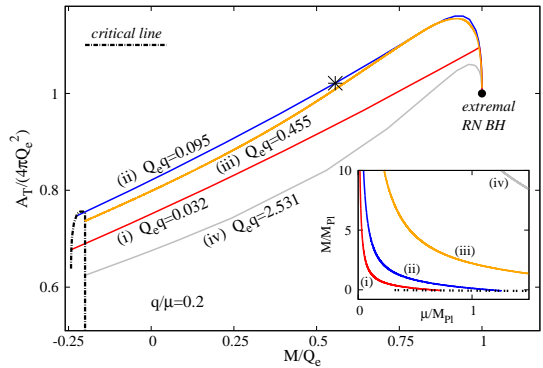}  
\caption{
The scaled area of the throat $vs.$ the scaled ADM mass is shown for
families of wormhole solutions.
The curves starts from the extremal Reissner-Nordstr\"om black hole and 
end in a critical line where the spatial localization of the spinors is lost.
}
\label{domain}
\end{figure}
 
 \begin{figure}[h!]
\begin{center}
\includegraphics[trim=50 0 50 0, clip,width=0.16\textwidth]{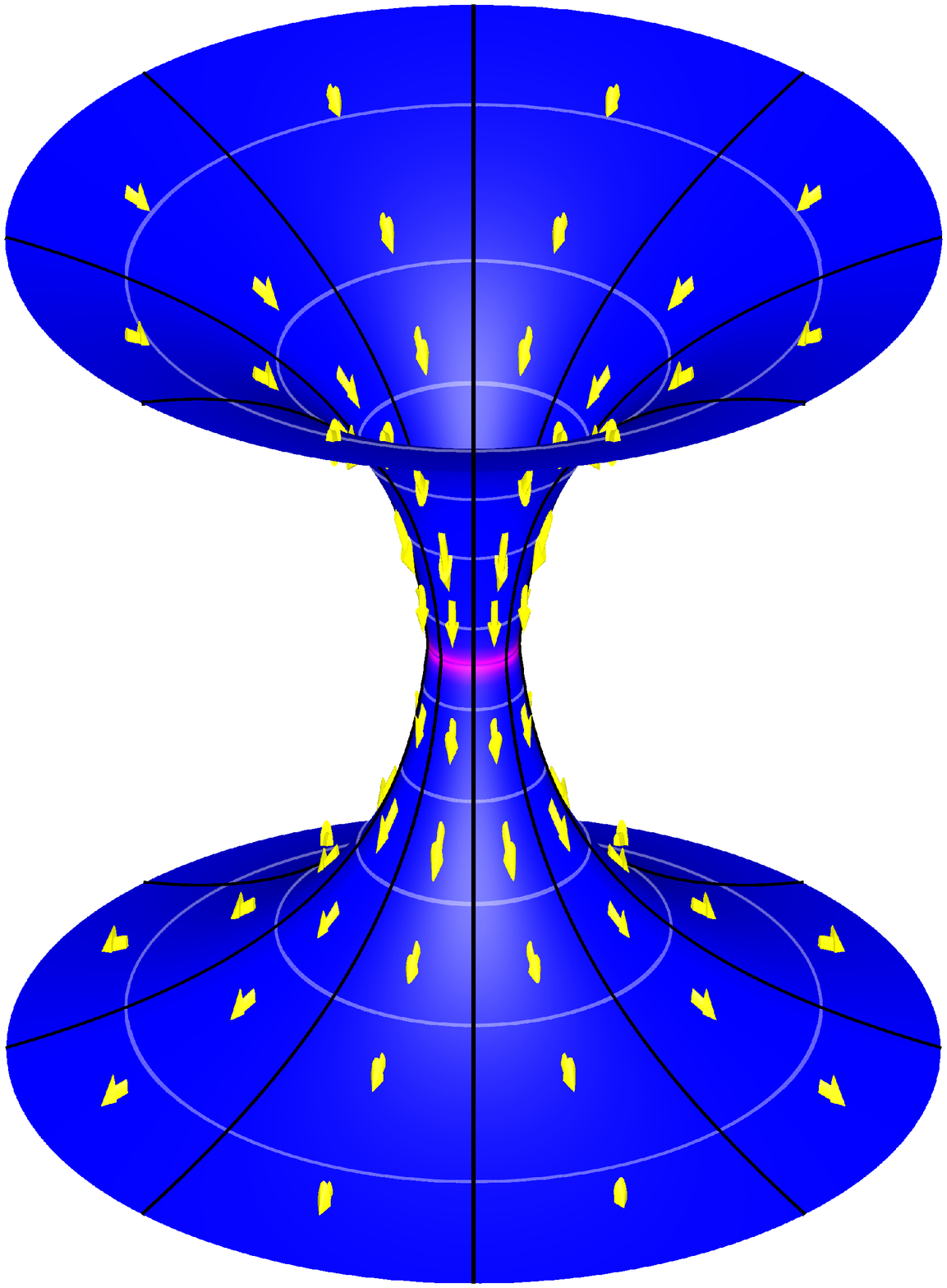}
\includegraphics[trim=50 0 50 0, clip,width=0.16\textwidth]{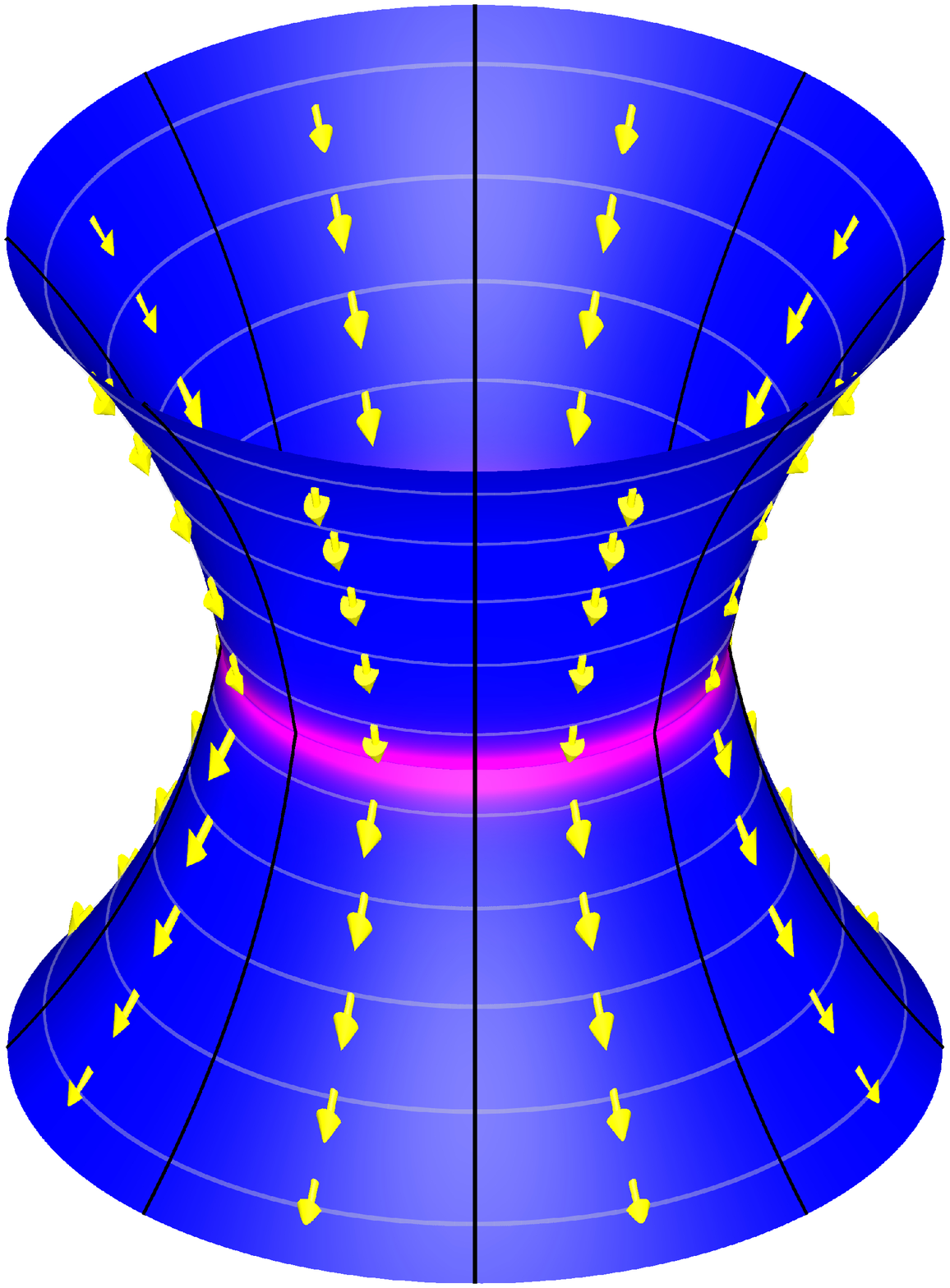} 
\includegraphics[trim=50 -20 50 50, clip,width=0.15\textwidth]{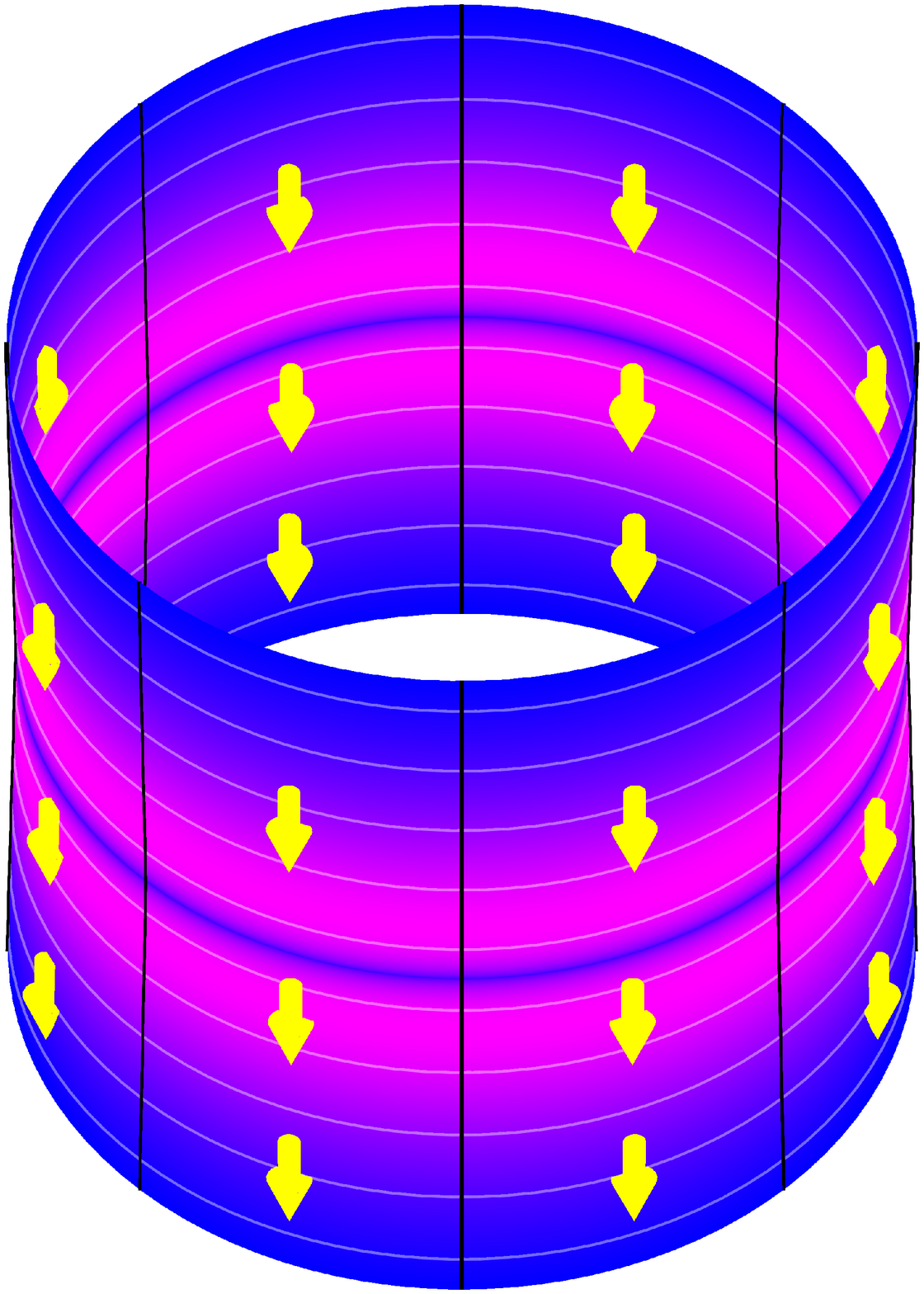} 
\includegraphics[width=0.40\textwidth]{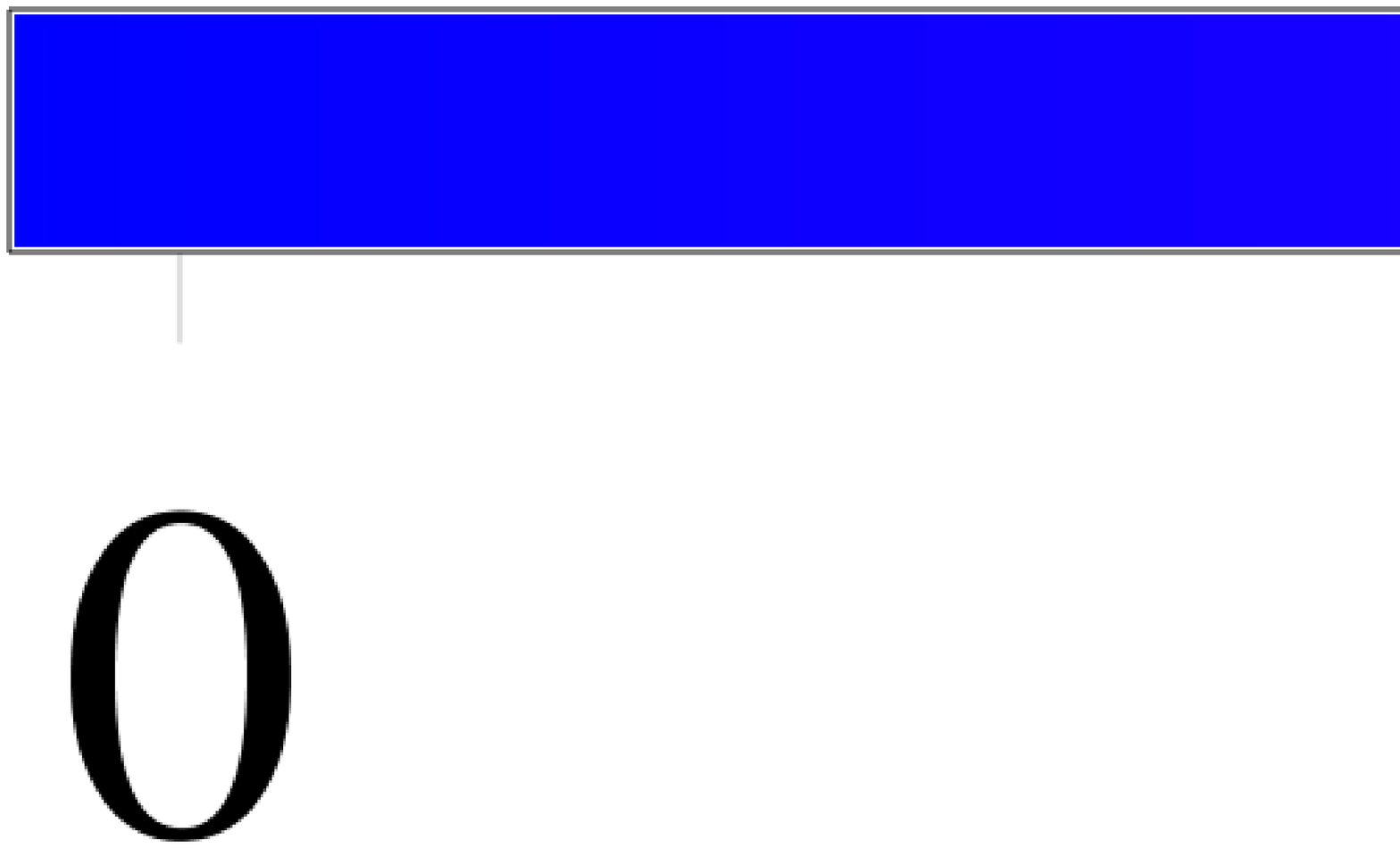}
\caption{A sequence of isometric embeddings, 
with a zoomed throat region,
is shown for the wormhole solution in Fig. 2.
The color map shows the absolute value of the  Noether charge
density.
}
 \label{embedding}
 \end{center}
 \end{figure}
 
A complementary picture is shown in 
the inset of Fig. 3, with the quantities given in Planck units 
 (with the one particle condition 
 imposed for each spinor).
 When the mass $\mu$ of the spinors is made smaller and smaller, 
the solutions get arbitrarily close to extremal RN BHs,
 while the ADM mass appears to increase without bounds  
(note that, since the product $Q_e q$ is constant
 along the colored lines in Fig. 2, 
$Q_e$ and $q$ behaves in this limit as $M$ and $\mu$, respectively).
 On the other hand, the largest values found for $\mu$ are of order $10^2 M_{Pl}$, 
being approached at the critical line.

 Essential for the existence of the WH solution 
 is the violation of the null energy condition
 $ T_{\mu\nu} n^\mu n^\nu \geq 0$, 
 for any null vector field $n^\mu$ \cite{Visser}.
The violation of this condition is displayed 
in Fig. 2, 
with 
$T_r^r-T_t^t  < 0$.
 The  isometric embedding of the same WH solution  
is shown in Fig. 4, where the $\theta=\pi/2$ plane is considered.
The (absolute value of the) Noether charge density is also plotted there as a colour map
(note that the maximal value of this quantity is approached outside the throat).

\noindent{\textbf{~~~EDM WHs and entanglement.--~}}
In addition, 
the `smooth', symmetric WHs have 
the Dirac fields at each side of the throat entangled in a particular way. 
Let us
introduce two observers (Alice and Bob), which live in the asymptotically flat regions, 
where the solutions are approximately those of
the flat space. 
Alice (at $r\to\infty$) sees the fermions in the state 
$\Psi^A_{\boldsymbol{\epsilon}} =\left|\omega,\kappa\right>$, 
while Bob (at $r\to-\infty$) sees the fermions in a state with opposite numbers, 
$\Psi^B_{\boldsymbol{\epsilon}} =\left|-\omega,-\kappa\right>$. 
The full asymptotic states will belong to the product of Alice and Bob Hilbert spaces,
with
$\Psi_{\boldsymbol{\epsilon}} ({|r|\to\infty}) 
=\Psi^A_{\boldsymbol{\epsilon}}\otimes\Psi_{\boldsymbol{\epsilon}}^B
=\left|\omega,\kappa\right>\otimes\left|-\omega,-\kappa\right>.$
 This corresponds to an entangled particle/antiparticle state of opposite chiralities  \cite{Cariglia:2018rhw}. 
The WHs cannot be `smooth' unless the fermions are entangled in such a way.  
Also, since the electric flux smoothly enters the throat on one side and exits on the other,
 Bob observes the opposite electric flux 
and also measures opposite  charges with respect to Alice (their frames being flipped). 
 
\noindent{\textbf{~~~Conclusions.--~}}
All known examples of 
traversable WHs with (classic)
bosonic fields
require 
some  
exotic matter and/or non-standard Lagrangians. 
However,  the results in this work show that the situation changes 
for a fermionic matter content. 
 WH solutions were found in the (standard) EDM theory, without introducing  
extra-matter in the bulk or at the throat,
providing
an explicit realization of Wheeler's idea of
``{\it electric charge without charge}'' \cite{Wheeler}.
  For the WHs to be `smooth', the presence of a total electric charge is crucial,
	while
to be traversable, the mass-charge ratio has to be smaller than one.

A semiclassical approach has been used, in which case the Dirac-Maxwell and 
Einstein equations are coupled,
the fermionic matter being treated as a quantum wave function,
a  treatment which may provide a reasonable
 approximation under certain conditions  \cite{ArmendarizPicon:2003qk}.
However, 
we expect such configurations to exist as well
 in a more complete setting, with fully quantized matter fields   \cite{note4},
as suggested by the results in \cite{Maldacena:2018gjk}.

 Also, although
we considered a simple toy model with two localized fermions,
this study can be extended to states
with an arbitrary number of fermions,
which would enhance the
size of quantum effects,
 while retaining the simplifications 
offered by spherical symmetry 
\cite{Finster:1998ws},
\cite{Leith:2020jqw}.
EDM WHs with a {\it single} spinor should also exist,  
possessing an intrinsic angular momentum \cite{Herdeiro:2019mbz}.
Generalizations of such WH solutions  for the full matter content of the Standard Model
are also likely to exist.
 
\noindent{\textit{\textbf{~~~Acknowledgements.--~}}}
J.L.B.S. gratefully acknowledges support from DFG Project No. BL 1553,
DFG Research Training Group 1620  \textit{Models of Gravity}.
The  work of E.R.  is  supported  by  the Center  for  Research  and  Development  in  Mathematics  and  Applications  (CIDMA)  
through  the Portuguese Foundation for Science and Technology (FCT - Fundacao para a Ci\^encia e a Tecnologia), 
references UIDB/04106/2020 and UIDP/04106/2020, and by national funds (OE), through FCT, I.P., 
in the scope of the framework contract foreseen in the numbers 4, 5 and 6 of the article 23,of the Decree-Law 57/2016, of August 29, changed by Law 57/2017, of July 19.  
We acknowledge support  from  the  projects  PTDC/FIS-OUT/28407/2017,  CERN/FIS-PAR/0027/2019 and PTDC/FIS-AST/3041/2020. This work has further been supported by the European Union’s Horizon 2020 research and innovation (RISE) programme H2020-MSCA-RISE-2017 Grant No. FunFiCO-777740.  
The authors would like to acknowledge networking support by the COST Action CA16104.



\begin{thebibliography}{99}
\bibitem{note3}
As with solitons \cite{Finster:1998ux},
WHs with excited Dirac spinors may exist as well, but here we consider fundamental solutions only.
Also, all numerical results have $\kappa=1$ (for $r>0$),
 although a (qualitatively) similar picture has been found for $\kappa=-1$.
\end{thebibliography}
\end{document}